\documentclass[fleqn,usenatbib]{mnras}

\usepackage{newtxtext,newtxmath}

\usepackage[T1]{fontenc}
\usepackage{ae,aecompl}
\usepackage{bigints}

\usepackage{graphicx}	
\usepackage{amsmath}	
\usepackage{amsmath, epsfig,natbib}
\usepackage{multicol}
\usepackage{color,ulem}
\usepackage{newtxtext,newtxmath}
\definecolor{webgreen}{rgb}{0,.5,0}
\definecolor{webbrown}{rgb}{.6,0,0}
\usepackage{subcaption}
\usepackage{physics}

\newcommand{\muMax}{\Delta \mu_{\max}}

\newcommand{\rbar}{$R_{\rm bar} \ $}

\title [Redshift evolution of bar-driven dark gaps] 
{Evolution of bar-induced dark gaps in galaxy discs: evidence of strong bar-driven effects already at $z > 2$}
\author[S. Chattopadhyay et al.]
	{Susnata Chattopadhyay,$^{1}$\thanks{E-mail: sc22ms077@iiserkol.ac.in}
     Soumavo Ghosh, $^{2}$ \thanks{E-mail: soumavo@iiti.ac.in}
     Dimitri A. Gadotti, $^{3}$ 
     Zoe A. Le Conte, $^{3}$ 
     Taehyun Kim, $^{4}$ 
     \newauthor
     Virginia Cuomo, $^{5}$
     Camila de S\'{a}-Freitas $^{6}$ and
     E. Athanassoula $^{7}$\\
$^1$ Department of Physical Sciences, Indian Institute of Science Education and Research Kolkata, Mohanpur, India - 741246\\
$^2$ Department of Astronomy, Astrophysics and Space Engineering, Indian Institute of Technology Indore, India - 453552\\
$^3$ Centre for Extragalactic Astronomy, Department of Physics, Durham University, South Road, Durham DH1 3LE, UK\\
$^4$ Department of Astronomy, Yonsei University, 50 Yonsei-ro, Seodaemun-gu, Seoul 03722, Republic of Korea. \\
$^5$ Departamento de Astronom\'{i}a, Universidad de La Serena, Avenida Ra\'{u}l Bitr\'{a}n 1305, La Serena, Chile\\
 $^6$ European Southern Observatory, Alonso de C\'{o}rdova 3107 Vitacura, Casilla 19001 Santiago de Chile\\
$^7$ Aix Marseille Univ, CNRS, CNES, LAM, Jardin du Pharo, 58 Boulevard Charles Livon, 13007 Marseille, France
}

 \date{Accepted XXX. Received XXX; in original form XXX}

\pubyear{2026}

\begin{document}

\nocite{*}
\label{firstpage}
\pagerange{\pageref{firstpage}--\pageref{lastpage}}
\maketitle
\begin{abstract} 
The properties of stellar bars play a crucial role in determining the bar-driven secular evolution in disc galaxies. However, a systematic observational study of the evolution of several bar properties (such as strength and length) across cosmic time is largely missing. In this paper, using a sample of $625$ barred galaxies, taken from SDSS, HST COSMOS, and JWST CEERS surveys, we systematically investigate the evolution of bar properties over redshifts ($0.02 \lesssim  z < 3$) by making a novel usage of dark gap (preferential light deficit along the bar minor axis) properties as a proxy for bar properties. We show that the dark gap strength ($\muMax$) exhibits a weak evolution, increasing from higher redshifts ($z \sim 2.5$) and slightly declining towards lower redshifts ($z < 0.05$). Conversely, the extent of dark gaps ($R_{\rm DG}, R_{\rm dark}$; normalised by bar length) decreases moderately from $z \geq 1.4$ and remains constant thereafter. Our results suggest that bar formation and the initial rapid growth phase occur before $z \sim 3$, followed by mild growth towards lower redshifts. We also find $R_{\rm dark}$ to be a better proxy (as compared to $R_{\rm DG}$) for estimating bar length, supporting earlier theoretical studies. Furthermore, the $\muMax$ shows a weak but statistically significant correlation with bar-to-total light ratio (Bar/T) and bar ellipticity ($\epsilon_{\rm bar}$). Studies of the redshift evolution of bar properties over such an extensive redshift range as done here are instrumental in constraining the bar-driven evolution at early cosmic times.
\end{abstract}


\section{Introduction}
\label{sec:Intro}

Stellar bars are one of the most common non-axisymmetric structures in disc galaxies in the local Universe. The fraction of disc galaxies in the local universe, hosting a prominent stellar bar at their central region, varies with the wavelength. For example, the bar fraction is measured up to $\sim 50$ percent in the optical wavelengths, while in infrared wavelengths, the bar fraction increases up to $\sim 70$ percent of the whole disc galaxy population in the local Universe \citep[e.g. see][]{Eskridgeetal2000, Menedesetal2007, NairandAbraham2010, Mastersetal2011, Butaetal2015, Kruketal2017}. The bar fraction as well as the bar properties are also shown to vary with stellar mass and Hubble type \citep[e.g. see][]{Kormdendy1979, Aguerrietal2005, MarinovaandJogee2007, Gadotti2011, Aguerrietal2009, Butaetal2010, NairandAbraham2010, Barwayetal2011, Erwin2018}. Bars are observationally found at higher redshift ($z \sim 1$) as well \citep[e.g. see][]{Shethetal2008,Melvinetal2014,Simmonsetal2014,EuclidCollab2025}. In addition, an effort to extrapolate the ages of old nearby bars and the evolution of bar fraction over time from a local universe perspective, has also been made \citep[e.g. see][]{camila2025} using samples from TIMER survey \citep{Gadotti2019}.
Furthermore, recent JWST observations revealed prominent bars up to redshift $z \sim 4$ \citep[e.g. see][]{Guoetal2022, Tsukui2023, Smailetal2023, Costantinetal2023, LeConteetal2024, Guoetal2025, Geron2025, Amvrosiadis2025b, LeConteetal2026}.   
On the other hand, recent theoretical studies have shown that disc galaxies with a massive kinematically-hot (e.g. higher velocity dispersion) thick disc, a dynamical situation mimicking the high redshift galaxies, can still form conspicuous bar and boxy/peanut bulges \citep{Ghoshetal2023a,Ghoshetal2023b}. In addition, recent state-of-the-art cosmological simulations showed that bar formation already starts beyond $z \sim 1$ \citep[e.g. see][]{Kraljicetal2012,Fragkoudietal2021,Rosas-Guevaraetal2022,Fragkoudietal2024}. However, the question remains - at what redshift do bars start to form? And how rapidly do their properties evolve with time?
\par
Understanding the bar formation epoch and evolution of their properties (strength, length and pattern speed) with redshift is extremely crucial as the dynamics and secular evolution of disc galaxies depend on the properties of the bar \citep{Gadotti2020}. Bars play a pivotal role in redistributing stars and reshaping metallicity distributions by radial migration \citep[e.g.][]{DiMatteoetal2013,Kubryketal2013,Halleetal2015,Khoperskovetal2020,Haywoodetal2024}, driving vertical bulk motions \citep{Monarietal2015,Khoperskovetal2019,Khachaturyantsetal2022}, reshaping the composition of the stellar component within the bar region \citep[e.g. see][]{Seideletal2016}, exciting ridge-like features in the phase-space \citep[e.g. see][]{Dehnen2000,Fragkoudietal2019,Tricketal2021}, and funnelling gas in the inner region of galaxies; thus facilitating starbursts and formation of nuclear discs \citep[e.g.][]{Shlosmanetal1990,Shethetal2005}. For a detailed exposition of bar-driven secular evolution in disc galaxies, the reader is referred to \citet{Kormdendy1979} and \citet{kormendyKennicutt2004}. Therefore, it is essential to gain a complete knowledge of the evolution of bar properties as a function of redshift. 
\par
As barred galaxies evolve, stellar bars grow in strength and extent with time - continuously trapping more stars onto the radially-elongated $x_1$ orbits; thus serving as a backbone for the resulting bar structure \citep[e.g. see][]{ContopoulosandGrosbol1989,Athanassoula2003,BinneyTremaine2008}. As a result, an initial azimuthally smooth light profile transforms into a radially bright light profile along the bar major axis, thereby causing a prominent deficit of light along the bar minor axis \citep[e.g. see][]{ GadottiSouza2003, Kimetal2016,Buta2017,Aguerrietal2023,Ghoshetal2024, Kimetal2025}. This light deficit along the bar minor axis is coined as `dark gap' \citep{Buta2017}. Previous $N$-body models showed that dark gaps and the stellar bar evolve in tandem \citep[e.g. see][]{Kimetal2016,GhoshDiMatteo2024,Ghoshetal2024}, and the properties of dark gaps (e.g. strength, extent) are strongly correlated with the properties of stellar bars \citep{Aguerrietal2023,GhoshDiMatteo2024,Ghoshetal2024}. In addition, \citet{Kimetal2016} showed that the strength of the dark gap is strongly correlated with the bar size and to the bar-to-total light ratio, from a sample of barred galaxies from the \textit{Spitzer} Survey of Stellar Structure in Galaxies (S$^4$G). 
\par
Past observational efforts linked the dark gaps in barred galaxies to photometric signatures of different bar resonances. \citet{Buta2017} argued that the dark gaps are the photometric signature of bar co-rotation. More recent studies by \citet{Krishnaraoetal2022} and \citet{Aguerrietal2023}, using a sample of MaNGA barred galaxies (supplemented by an $N$-body model of a barred galaxy) and a sample of barred galaxies from CALIFA, ESO/MUSE, ESO/NTT observations, suggested that the locations of the dark gaps are associated with the 4:1 ultra-harmonic resonance of the bar. However, a systematic study by \citet{Ghoshetal2024}, using a suite of $N$-body models (with varying structural parameters), demonstrated that the location of dark gaps is not a universal proxy for the bar resonances (co-rotation, Inner Lindblad resonance and 4:1 ultra-harmonic resonance), in contrast with earlier studies. Furthermore, by comparing the locations of dark gap and resonance radii, \citet{Kimetal2025} found that, among various galaxy types, only certain morphological types exhibit dark gaps that align with specific resonances. While the dark gaps may not be a universal proxy for the bar resonances (or equivalently, proxy to measure the bar pattern speed), it is well understood that the properties of dark gaps (e.g. strength and extent) serve as an excellent proxy for the strength and length of bar \citep[e.g. see][]{Aguerrietal2023,Ghoshetal2024}.
\par
Despite the challenges involved, studying bar properties at high redshifts is becoming important for understanding their overall evolution. In this work, we aim to carry out a study of the evolution of bar properties across a wide redshift range (up to $z \sim 3$) by making novel usage of the dark gap as a proxy for the bar properties. Using bar samples from the SDSS, HST COSMOS and JWST CEERS surveys (introduced later), we systematically measure the properties of dark gaps (strength and extent) and also investigate their variation with redshift. We further study the dependence (if any) of dark gaps on other photometric properties of the bar. Lastly, we examine the effects on dark gap properties, of possible biases due to projection effects, choice of photometric band effects, and choice of survey.
\par
The rest of the paper is organized as follows: Sec.~\ref{sec:bar_sample_main} provides a brief description of the bar samples selected from the SDSS, HST COSMOS and JWST CEERS surveys. Sec.~\ref{sec:redshift_darkgap} provides the results pertaining to redshift evolution of dark gap properties as well as their connection with other structural properties of host galaxies. Sec.~\ref{sec:discussion} discusses the implications of this work while Sec.~\ref{sec:summary} summarizes the main findings of this work.

\section{Sample of barred galaxies}
\label{sec:bar_sample_main}

In this work, we make use of the photometric data of barred galaxies obtained from three major astronomical surveys, namely the Sloan Digital Sky Survey  \citep[SDSS,][]{York2000, Abazajian2004b}, the Cosmic Evolution Survey \citep[COSMOS, ][]{scoville2007, koekemoer2007} from the Hubble Space Telescope (HST), and the Cosmic Evolution Early Release Science \citep[CEERS, ][]{FinkelsteinEtal2025} from the James Webb Space Telescope. The sample of barred galaxies from SDSS is taken from \citet{Gadotti2009} whereas the samples from HST COSMOS and JWST CEERS surveys are taken from \citet{kim2021} and \citet{LeConteetal2026}, respectively. For the sake of completeness, we briefly describe the properties of the samples of barred galaxies and the associated selection criteria. For further details, the reader is referred to the aforementioned papers. 

\subsection{SDSS bar sample}
\label{sec:sdss_bar}
SDSS is a wide-field ultra-violet to  near-infrared imaging and spectroscopic survey conducted at Apache Point Observatory (APO),  New Mexico using a dedicated 2.5m f/5 telescope with a field of view of 3$^{\circ}$. SDSS provides observation of photometric properties of galaxies in a total of five different wave-bands, namely, $\qty( u, \ g, \ r, \ i, \ z)$ centered at  $\qty(357.3,\ 472.3, \ 620.2, \ 752.2, \ 891.2)$ nm \footnote{at an \texttt{AIRMASS} = 1.3} at magnitude limits of $\qty(22.0, \ 22.2, \ 22.2, \ 21.3, \ 20.5)$ mag respectively \citep{Abazajian2004b}. The size of each pixel is 24 $\mu$m, which corresponds to 0.396$''$ in the sky.
\par 
The barred galaxies from SDSS (based on $i$-band photometry) are taken from \citet{Gadotti2009} which provides a comprehensive study of properties of bars in (nearly) face-on galaxies \citep{Gadotti2011}. They are part of SDSS Data Release 2 (DR2) and have galaxies in redshift range $0.02 \lesssim z \leq 0.07$. After removing galaxies with foreground star contamination and highly disturbed morphology, our final selected SDSS barred sample consists of 264 face-on galaxies.

\subsection{HST COSMOS bar sample}
\label{sec:hst_bar}

COSMOS is one of the largest surveys conducted by the HST. It is imaged in an equatorial field with single-orbit F814W (I-band $\sim 807.3$ nm ) \footnote{\url{https://cosmos.astro.caltech.edu/page/filterset}} exposures to a limiting magnitude of  $I_{AB} \simeq$ 28 mag. With very high resolution ($0.03''$ pixels) and sensitivity (FWHM of $0.09''$) along with a FOV (field-of-view) of 2 deg$^2$, this survey has enabled us to resolve morphologies of several hundred thousand galaxies; thereby enabling a thorough probe of the evolution of AGNs, dark matter and galaxies ($\sim 2 \times 10^6$ objects) within their cosmic surroundings \citep[e.g. see][]{scoville2007,koekemoer2007}. 
\par
In this work, the barred galaxies from the HST COSMOS survey are taken from \citet{kim2021}\footnote{ \cite{kim2021} used the drizzled data from \url{https://irsa.ipac.caltech.edu/data/COSMOS/images/acs_mosaic_2.0/ACS_INFO_2.0.txt}} which provides a comprehensive study of bars in the context of the COSMOS fields. The selected sample of barred galaxies encompasses a redshift range of $0.2 \lesssim z  \lesssim 0.83$. After excluding galaxies with foreground star contamination, highly disturbed morphologies and divergent brightness profiles (similar to the procedure applied to our SDSS bar sample), the final HST COSMOS barred galaxy sample consists of 299 galaxies.

\subsection{JWST CEERS bar sample}
\label{sec:jwst_bar}

The CEERS survey is one of the flagship programs being conducted with the JWST. The Near-Infrared Camera (NIRCam)\footnote{For documentation about JWST and its instruments including NIRCam, visit  \url{https://jwst-docs.stsci.edu/}} is one of JWST’s primary cameras, operating in the wavelength band from 600 to 5000 nm \citep[e.g. see][]{Rieke2023}. 
\par
In this work, the barred galaxies from the JWST CEERS survey are taken from \citet{LeConteetal2026} which provides a sample of barred galaxies, detected via a 3-step optimization procedure, namely, an initial elliptical isophotal fitting, a fixed-centre isophote-fitting, and finally, removal of galaxies with inclination $i > 60^{\circ}$ \citep[for further details, see][]{LeConteetal2024}. After removing samples with  disturbed morphology, our final selected JWST CEERS barred sample consists of 27 galaxies from the short wavelength channel (hereafter, CEERS F200W sample - it is obtained at $0.6 - 2.3 \ \mu $m) and 35 galaxies from the long wavelength channel (hereafter, CEERS F444W \footnote{Note for galaxies between $1 < z < 2$ they are in F356W filter, and for $z > 2$ they are in F444W.} sample - corresponding to $2.4 - 5.0 \ \mu $m wavelengths). The short and long wavelength channels have nominal pixel scales of $0.031''$ and $0.063''$ per pixel respectively. This combined sample primarily encompasses a redshift range of $1.0 \leq z < 3$ (with only one galaxy having redshift $z \sim 3.2$). Our JWST CEERS sample size is small as compared to the other two samples considered in this work, predominantly due to the challenges in detecting faint and lower mass barred galaxies at these high redshifts, thus rendering our sample to be limited with relatively large error bars (as discussed in Sec.\ref{sec:redshift_darkgap}).
\par
\begin{figure}
    \centering
    \includegraphics[width=\linewidth]{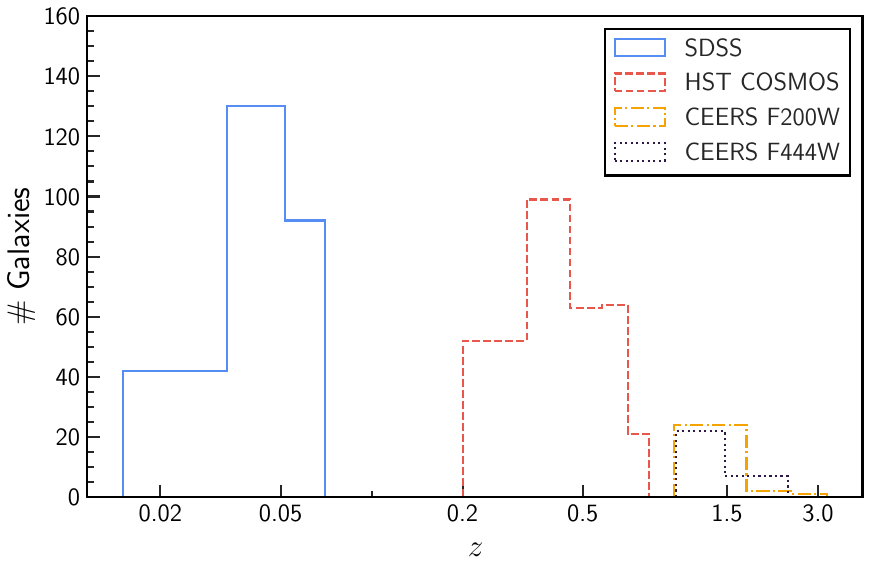}
    \caption{Redshift distribution of the  selected barred samples from the SDSS, HST COSMOS and JWST CEERS surveys, used in this work. For further details, see the text in Sec.~\ref{sec:bar_sample_main}.}
    \label{fig:z_samples}
\end{figure}

\begin{figure}
    \centering
    \includegraphics[width=\linewidth]{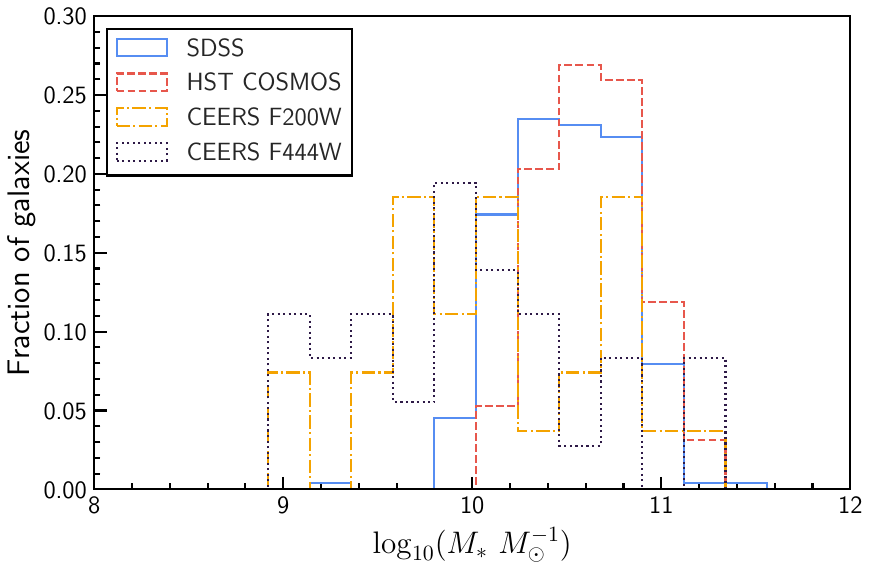}
    \caption{Stellar mass distribution of the  selected barred samples from the SDSS, HST COSMOS and JWST CEERS surveys, used in this work.}
    \label{fig:mass_samples}    
\end{figure}

 
In total, we used 625 barred galaxies taken from three different surveys. As shown in Fig.~\ref{fig:z_samples}, they span over a wide of range of redshift ($0.02 \lesssim z < 3$). This enables us to carry out a novel systematic study of the bar-driven dark gap properties over such a wide redshift range. The corresponding stellar mass distribution of our sample of galaxies from different surveys is shown in Fig. \ref{fig:mass_samples}. The JWST CEERS samples are measured to have stellar masses \citep[obtained from][]{LeConteetal2026} almost uniformly in the order of $10^9 - 10^{11.5} \ M_{\odot}$ (solar masses), while most of SDSS and HST COSMOS samples lie between $10^{10} - 10^{11} \ M_{\odot}$. The stellar masses of the galaxies in the SDSS sample are taken from \cite{Gadotti2009}, while that of the HST COSMOS samples belong to \cite{kim2021}. Throughout this paper, we assume the latest \texttt{PLANCK} flat $\Lambda$CDM cosmology model with $H_0$ = 67.36 km s$^{-1}$ Mpc$^{-1}$, $\Omega_{\mathrm{m}}$ = 0.3153 and $\Omega_{\Lambda}$ = 0.6847 \citep{PlanckCollab2020} to calculate the scale factors. We have used the calculator developed by \cite{Wright2006} for this purpose.

\section{Redshift evolution of bar-driven dark gaps}
\label{sec:redshift_darkgap}

\begin{figure*}
    \centering
    \includegraphics[width=\linewidth]{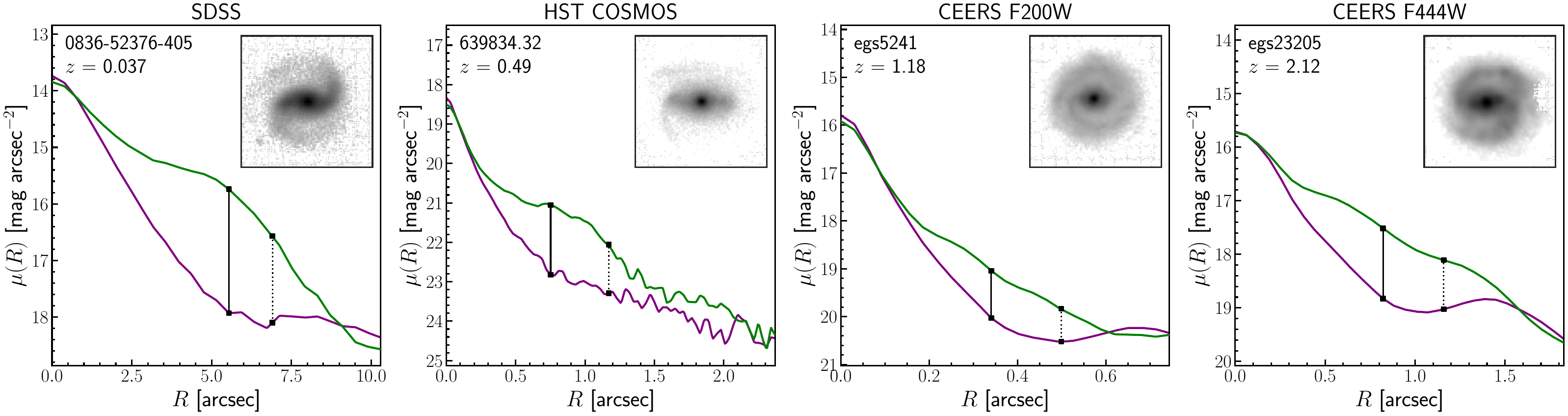}
    \caption{\textit{Dark gap in our sample of barred galaxies}: each of the panels corresponds to a barred galaxy (shown in inset) selected from the SDSS, HST COSMOS and JWST CEERS samples respectively, as well as the corresponding radial surface brightness profiles (with the bar aligned with the $x$-axis) along the bar major (green) and minor (purple) axes. The vertical \textit{solid} black line denotes the location of $R_{\rm DG}$, that is, the spatial location where $\Delta \mu (R)$ is maximum. While the vertical \textit{dotted} black line denotes the location of $R_{\rm dark}$, that is, the spatial location beyond $R_{\rm DG}$ where $\Delta \mu (R)$ drops to 70\% of its maximum value ($\muMax$). For further details, see the text.}
    \label{fig:darkgap_example1}
\end{figure*}
 
 Here, we investigate the evolution of dark gap properties using our selected sample of barred galaxies, chosen from the SDSS, HST COSMOS, and JWST CEERS surveys. To achieve this, we begin by constructing radial surface brightness profiles, $\mu(R)$,  along the bar major and minor axes. Before extracting these radial profiles, we follow the steps outlined below. 
 
 \begin{itemize}
     \item  First, we de-project the galaxy using the knowledge of the inclination angle ($i$) and imposing the flux conservation condition \citep[for further details, see e.g.][]{kim2021}. The SDSS samples are (almost) face-on \citep[with axial ratio $b/a \geq 0.9$ measured in $g$ band - 25 mag arcsec$^{-2}$ isophote; for further details see][]{Gadotti2009}, so the de-projection procedure is not required. The values of $i$ for the HST COSMOS sample are taken from \citet{kim2021}. However, for the JWST CEERS sample, given the larger uncertainties of $i$, we refrain from doing the de-projection for this subsample. As shown in Appendix~\ref{appen:item_A}, we found that the de-projection procedure does not change the $\muMax$ values appreciably. For further details of the effect of the de-projection on determining the dark gap properties, the reader is referred to Appendix~\ref{appen:item_A}. 
     \item  Then, using the bar PA (position angle), obtained from the photometric analysis, we rotate the bar in such a way that the bar major axis always coincides with the horizontal $x$-axis. 
     \item Finally, we put a slit of width $\delta$ along the bar major and minor axes, and calculate the corresponding (smoothed) radial surface brightness profiles. For SDSS, we set $\delta = 5$ pixels and for HST COSMOS and JWST CEERS, $\delta = 3$ pixels. We mention that the equivalent widths of the slits are  $1.98^{''}$ for SDSS, $0.09^{''}$ for HST COSMOS, $0.09^{''}$ for CEERS F200W, and $0.19^{''}$ for the CEERS F444W samples. The values of $\delta$ are set so as to get a smooth radial profile along the bar major and minor axes.
    \end{itemize} 

Fig.~\ref{fig:darkgap_example1} shows an example of 4 barred galaxies selected from our SDSS, HST COSMOS, and JWST CEERS surveys. A mere visual inspection reveals that for each of these cases, a prominent dark gap is associated with the bar.
In the following subsections, we quantify the (average) values of the dark gap properties (strength and extent) in different redshift bins, and further investigate their correlation (if any) with other photometric properties of bars.

\subsection{Evolution of strength and extent of dark gaps}
\label{subsec:darkgap_properties_evolution}

\begin{figure}
    \centering
    \includegraphics[width=1\linewidth]{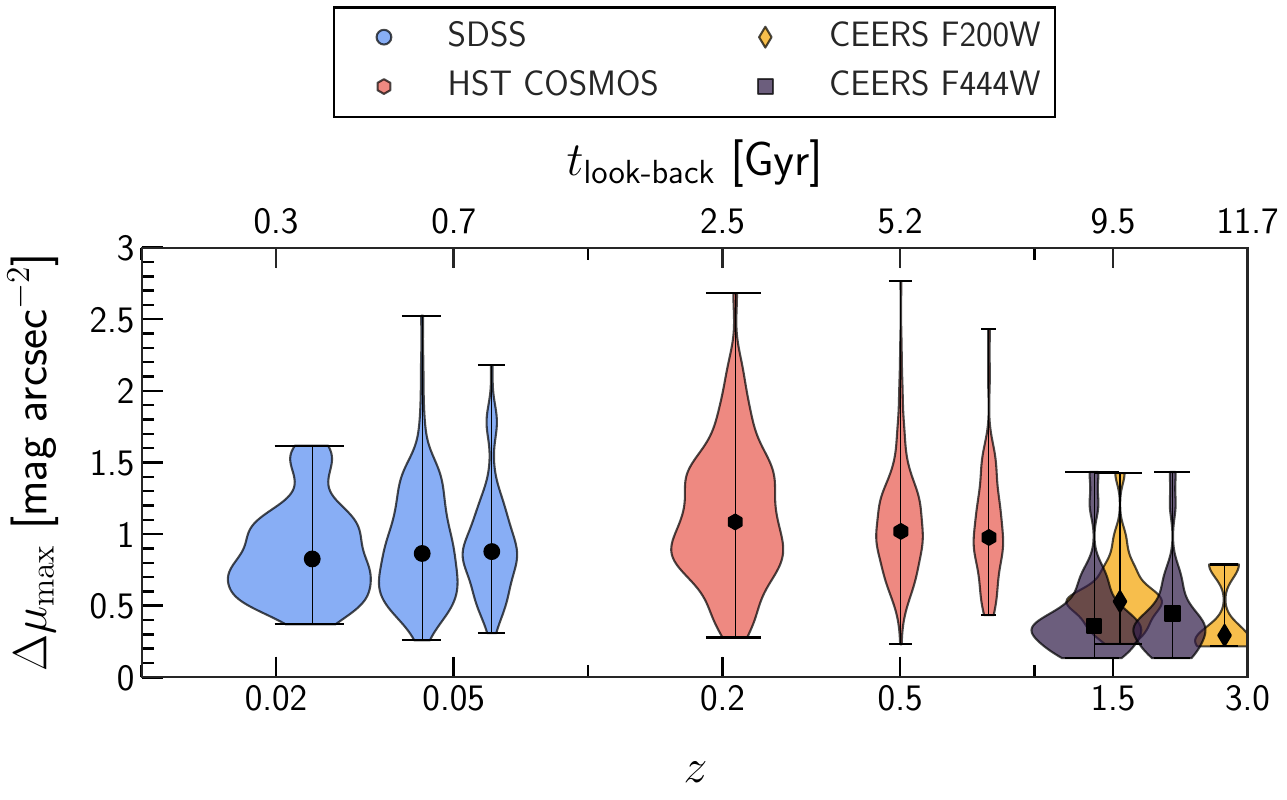}
    \caption{\textit{Redshift evolution of dark gap strength:} the distribution of dark gap strength, $\muMax$, calculated at different redshift bins, for our selected sample of barred galaxies from the SDSS, HST COSMOS, and JWST surveys (see the legend). The median values are indicated by the black markers. The width of the plots indicates the probability density of the distribution. The widths are not directly comparable, as each is scaled individually per sample to enhance the visibility. The solid black vertical lines denote the redshift bin centers, while the horizontal lines depict the entire extent of $\muMax$ values in that particular redshift bin. The axis on the top represents the corresponding look-back time, $t_{\rm look-back}$ (in units of Gyr). The median values of $\muMax$ remain almost constant in the intermediate redshift regime ($0.2 \leq  z \leq 1$) whereas a moderately rising trend is seen at the higher redshift regime ($z \sim 2.5$) and a slightly decreasing trend towards the low redshift regime ($z \sim 0.02$).}
    \label{fig:muMax_z}
\end{figure}

Following \citet{Kimetal2016} and \citet{Ghoshetal2024}, we define the dark gap strength as the maximum light deficit ($\Delta \mu_{\max}$) between the radial surface brightness profiles along the bar major and minor axes (see Fig.~\ref{fig:darkgap_example1}). By construction, it is a positive quantity. Furthermore, we define the dark gap radius, $R_{\rm DG}$ as the radial location where the maximum light deficit $\Delta \mu_{\max}$ occurs \citep[for details, see e.g.][]{Kimetal2016,Aguerrietal2023,Ghoshetal2024}. In other words, $\Delta \mu (R = R_{\rm DG}) \equiv \Delta \mu_{\max}$. In addition, \citet{Ghoshetal2024} introduced another measurement for the dark gap extent, namely,  $R_{\rm dark}$ which is defined as the radial location where the values of $\Delta \mu$ drop to 70 percent of its peak value ($\Delta \mu_{\max}$). Mathematically, $R_{\rm dark}$ can be expressed as 

\begin{equation}
R_{\rm dark} = \sup \{ R: \mu_{\rm minor} (R) - \mu_{\rm major} (R)  \geq  0.7 \muMax \} \,.
\end{equation}
\noindent The main difference between $R_{\rm DG}$ and $R_{\rm dark}$ is that one locates the peak of $\Delta \mu$ while the other extends beyond the peak location of $\Delta \mu$. For further details, see \citet{Ghoshetal2024}.

Next, using the procedures mentioned in Section ~\ref{sec:redshift_darkgap}, we first extracted the radial surface brightness profiles along the bar major and minor axes and then computed the value of $\muMax$ for all 625 barred galaxies chosen from three different surveys. The corresponding full distribution of $\muMax$ values in different redshift bins is shown in Fig.~\ref{fig:muMax_z}. We mention that each of these bar samples (taken from different surveys) has a finite redshift range. Therefore, within a given sample of barred galaxies (taken from a particular survey), we partition the redshift intervals in bins of equal sizes. This in turn, ensures a fair representation of galaxies across all redshifts.
As seen from Fig.~\ref{fig:muMax_z}, the average values of $\muMax$ display an intriguing profile with redshift. To elaborate, from the higher redshift regime ($z \sim 2.5$), $\muMax$ shows a moderately rising trend ($\sim 85\%$) whereas in the intermediate redshift range ($0.2 \leq  z \leq 1$), the values of $\muMax$ remain largely constant. Only towards the low-redshift regime ($z \sim 0.02$), the values of $\muMax$ hint a slight decrease with decreasing redshift. Furthermore, we checked that in the entire redshift range ($0.02 \lesssim z < 3$), the $\muMax$ values are changed only by around $50$ percent, and the dark gap strengths do not show any drastic evolution (at least, in the statistical sense) throughout the wide range of redshift chosen for this work. 
\par 
%
\begin{figure}
    \centering
    \includegraphics[width=1\linewidth]{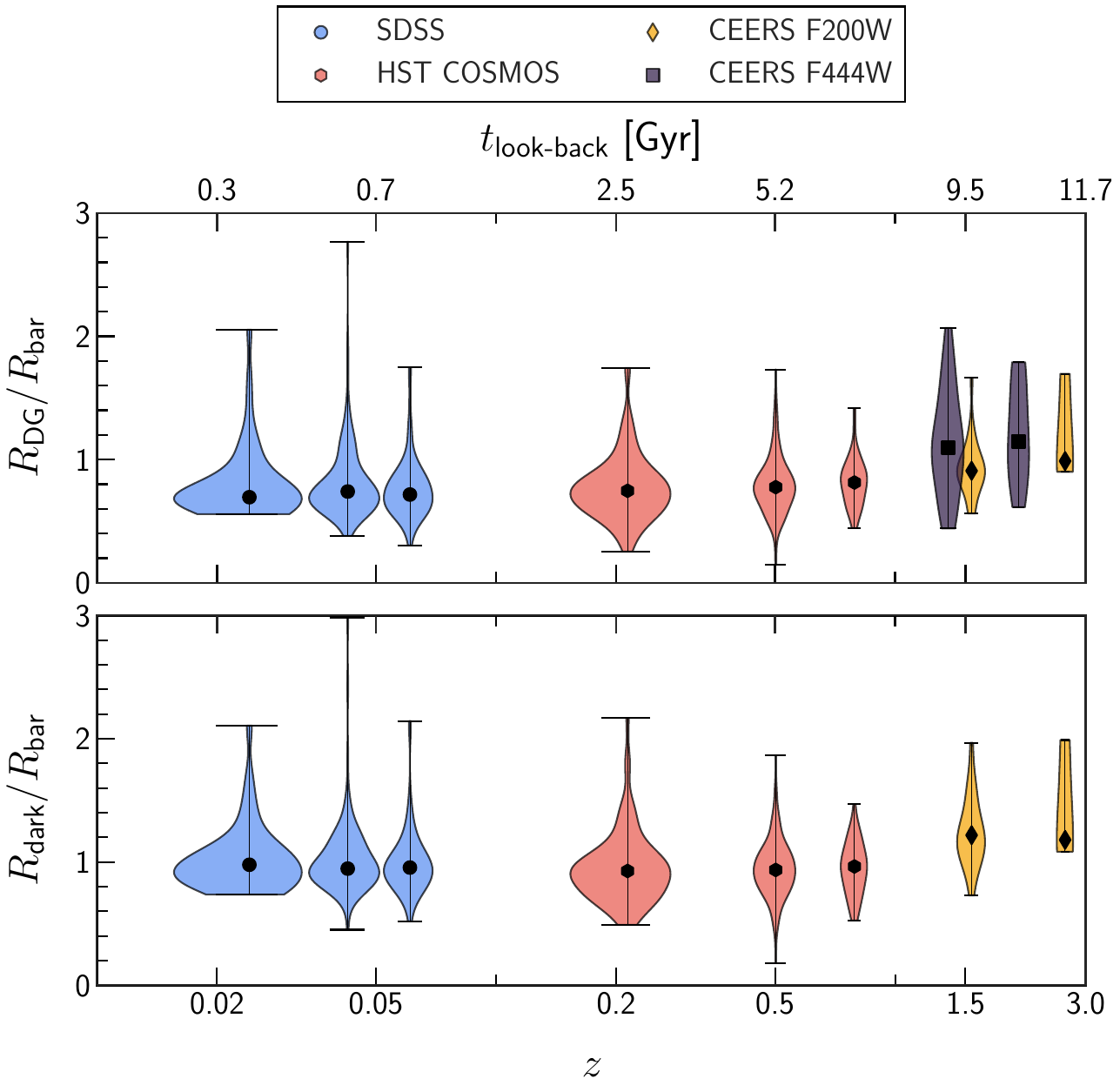}
    \caption{\textit{Redshift evolution of dark gap extents:} the distribution of dark gap extents, $R_{\rm DG}$ (top panel) and $R_{\rm dark}$ (bottom panel), each normalised by the \rbar, at different redshift bins, for our selected sample of barred galaxies from SDSS, HST COSMOS, and JWST surveys are shown here. The median values are indicated by the black markers. The width of the plots indicates the probability density of the distribution. The widths are not directly comparable, as each is scaled individually per sample to enhance the visibility. The solid black vertical lines denote the redshift bin centers, while the horizontal lines depict the entire extent of $R_{\rm DG}$ (or $R_{\rm dark}$) values in that particular redshift bin. The axis on the top represents the corresponding look-back time, $t_{\rm look-back}$ (in units of Gyr). The median values of (normalised) dark gap extents show a remarkably (almost) constant value for a wide range of redshift, decreasing only by $\sim 28$ percent from high redshift regime ($z \sim 2.5$) to the lower redshifts.}
    \label{fig:extents_r90_z}
\end{figure}

%
\par

Next, we investigate the redshift evolution of the extent of dark gaps from our  selected barred samples. First, we calculated the values of $R_{\rm DG}$ and $R_{\rm dark}$ (using the definitions mentioned in Sec.~\ref{sec:redshift_darkgap}) for all barred galaxies chosen for this work. We mention that for the CEERS F444W sample, accurate calculation of $R_{\rm dark}$ is not possible for most samples, primarily due to the presence of conspicuous spirals which contaminate the radial variation of $\Delta \mu$ produced by the bar, in addition to the limited resolution of images at high redshifts and long-wavelengths. As a result, we leave out the CEERS F444W sample in subsequent $R_{\rm dark}$ related analyses.  Furthermore, both $R_{\rm DG}$ and $R_{\rm dark}$ (in the other samples) are normalised by the bar length ($R_{\rm bar}$) for a uniform comparison among different galaxies with varied sizes. We mention that, in principle, one could also think of using $R_{90}$ (radius containing 90\% of the total galaxy luminosity), to normalise the dark gap extents. But, we avoided using it in our work due to its complex redshift artifacts. 
Moreover, the point spread function (PSF) itself blurs the light distribution. For example, the outer low surface-brightness emission of high-redshift galaxies may fall below the detection limit after PSF convolution, making the \textit{observed} light profile appear more centrally concentrated, and hence leading to an \textit{underestimation} of $R_{90}$. Additionally, since our sample is drawn from multiple surveys, the reliability of $R_{90}$ values are vastly different and thus do not provide a uniform comparison across the entire redshift range in our study ($0.02 \lesssim z < 3$). On the other hand, while using $R_{\rm bar}$ as the normalising factor, any redshift biases affecting the measure of $R_{\rm DG}$ or $R_{\rm dark}$ would be same as in the measurement of $R_{\rm bar}$, and those biases essentially get canceled out. The corresponding full distributions of  $R_{\rm DG}$ and $R_{\rm dark}$ (normalised by \rbar) at different redshift bins are shown in Fig.~\ref{fig:extents_r90_z}. We have used the same redshift binning as in Fig.~\ref{fig:muMax_z}. Even a mere visual inspection of Fig.~\ref{fig:extents_r90_z} reveals that both the metrics quantifying the (normalised) extent of dark gaps, namely, $R_{\rm DG}$ and $R_{\rm dark}$, remain remarkably constant over a wide redshift range. To elaborate, from the higher redshift regime ($z \sim 2.5$), the values of both $R_{\rm DG}$ and $R_{\rm dark}$ display a moderate decrease ($\sim 28$ \%), and thereafter in the intermediate ($0.2 \leq  z \leq 1$) and low redshift regime ($z < 0.05$), they remain largely constant (change by less than $10 \%$). Moreover, we checked that in the entire span of redshifts ($0.02 \lesssim z < 3$), the values of $R_{\rm DG}$ and $R_{\rm dark}$ change only by around $32\%$ and $17\%$ respectively. Hence, we can conclude that the dark gap extents also do not show any drastic evolution (at least, in the statistical sense).
\par
\begin{figure}
    \centering
    \includegraphics[width=0.95\linewidth]{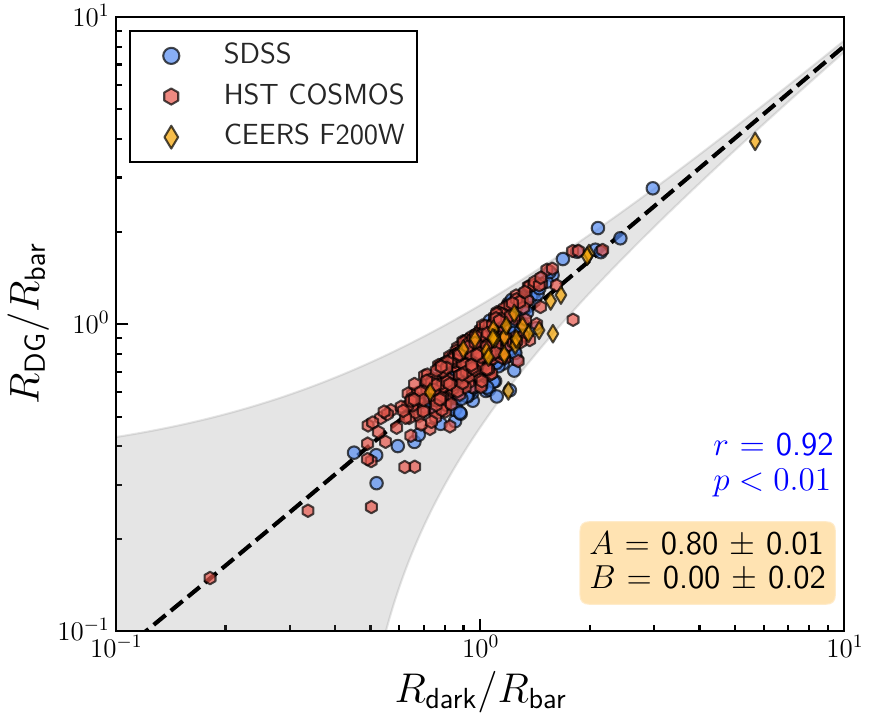}
    \caption{Correlation between the two dark gap extent estimators, $R_{\rm DG}$ and $R_{\rm dark}$, both normalised by \rbar, for our selected barred samples from the SDSS, HST COSMOS, and JWST surveys (see the legend). The correlation between $R_{\rm DG}$ and $R_{\rm dark}$ remain strong (Pearson correlation coefficient, $r > 0.75$) as well as statistically significant ($p-$value $< 0.01$).  The black dashed line denotes the best-fitting straight line of the form $Y = AX + B$, while the grey shaded region indicates a $3\sigma$ scatter around the best fit.}
    \label{fig:corr1_rdg_rDark}
   \end{figure}

 
Lastly, we examine if there is any correlation between the two metrics, namely, $R_{\rm DG}$ and $R_{\rm dark}$,  quantifying the extent of dark gaps. Using a suite of $N$-body bar models, \citet{Ghoshetal2024} showed that these two quantities are well correlated. However, it remains to be verified from observations (encompassing a wide range of redshift). Here, we pursue this. Fig.~\ref{fig:corr1_rdg_rDark} shows the distribution of all 625 galaxies from our selected samples in the $R_{\rm DG}-R_{\rm dark}$ plane. For a uniform comparison (among different galaxies with varied sizes), both the quantities are normalised by \rbar. To quantify the correlation, we computed the Pearson's correlation coefficient ($r$) as well as the corresponding $p$-value. As seen from Fig.~\ref{fig:corr1_rdg_rDark}, the correlation between $R_{\rm DG}$ and $R_{\rm dark}$ is found to be strong ($r > 0.75$) as well as statistically significant $(p<0.01)$, in agreement with the theoretical study of \citet{Ghoshetal2024}. To investigate further the universality of the correlation and statistical significance across different bar samples considered here, we computed $r$ and $p$ values, separately for each of these bar samples (see Appendix \ref{appen:item_C}). We found that, the values of $r$ are 0.92, 0.90 and 0.98 for the SDSS, HST COSMOS, and CEERS F200W samples respectively. Thus, the (strong) correlation between $R_{\rm DG}$ and $R_{\rm dark}$ remain unaffected by the choice of a specific survey.
In addition, we fit a straight line of the form $Y = AX+B$ (in Fig.~\ref{fig:corr1_rdg_rDark}) in order to investigate whether they follow any linear relation. We find $A = 0.80 \pm 0.01$ and $B = 0.00 \pm 0.02$, and almost all galaxies (except 4 outliers out of a total of 590 galaxies \footnote{Note that we have excluded CEERS F444W sample in Fig. \ref{fig:corr1_rdg_rDark}}) fall within a $3\sigma$ region around the best-fit straight line (see Fig.~\ref{fig:corr1_rdg_rDark}); thereby implying that these two quantities indeed follow a linear relation.
\par
%

\subsection{Correlation of dark gap properties with bar properties}
\label{sec:correlation_study}

%
\begin{figure*}
    \centering
    \begin{subfigure}{0.485\textwidth}
        \centering
        \includegraphics[width=0.74\linewidth]{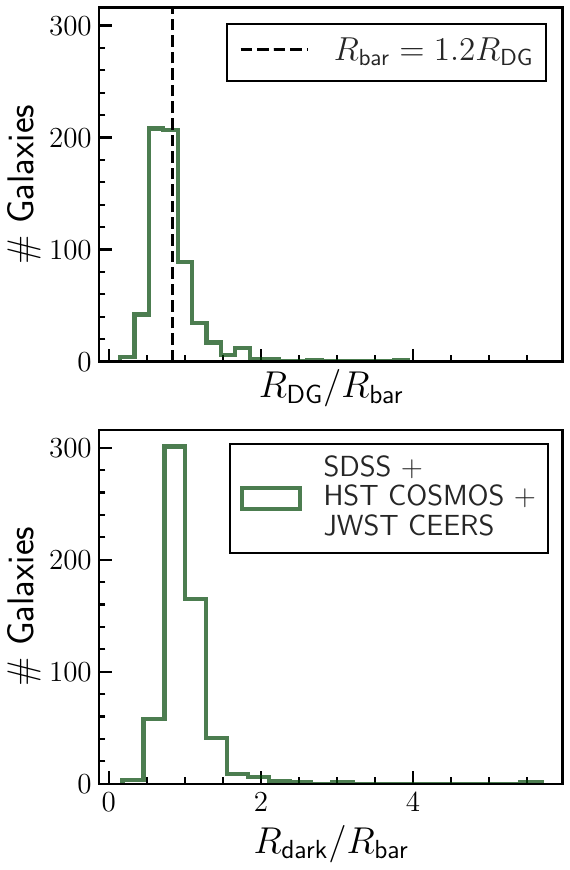}
    \end{subfigure}
    \hfill
    \begin{subfigure}{0.485\textwidth}
        \centering
        \includegraphics[width=0.8\linewidth]{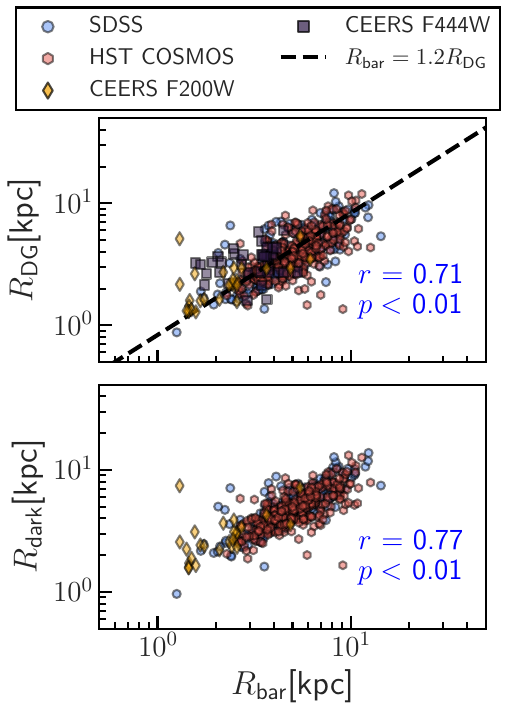}
    \end{subfigure}
    \caption{ \textit{Left panel}: Histograms of  $R_{\rm DG}/ R_{\rm bar}$ (top panel) and $R_{\rm dark}/ R_{\rm bar}$ (bottom panel), for all our combined barred samples from the SDSS, HST COSMOS, and JWST surveys (see the legend). \textit{Right panel:} Correlation between the bar length, $R_{\rm bar}$ and $R_{\rm DG}$ (top panel), and between $R_{\rm bar}$ and $R_{\rm dark}$ (bottom panel), all calculated (in kpc) for our selected barred samples from the SDSS, HST COSMOS, and JWST surveys (see the legend). In each case, the Pearson correlation coefficient, $r$ is calculated, and the corresponding values are quoted. $R_{\rm DG}$ is found to be slightly less correlated with bar length, as compared to $R_{\rm dark}$ which has a strong as well as statistically significant correlation ($r>0.75$ ,  $p-$value $< 0.01$) with \rbar, for our selected sample of barred galaxies.}
    \label{fig:corr2_rbar_rdg_rDark}
\end{figure*}

   
\begin{figure}
    \centering 
    \includegraphics[width=1\linewidth]{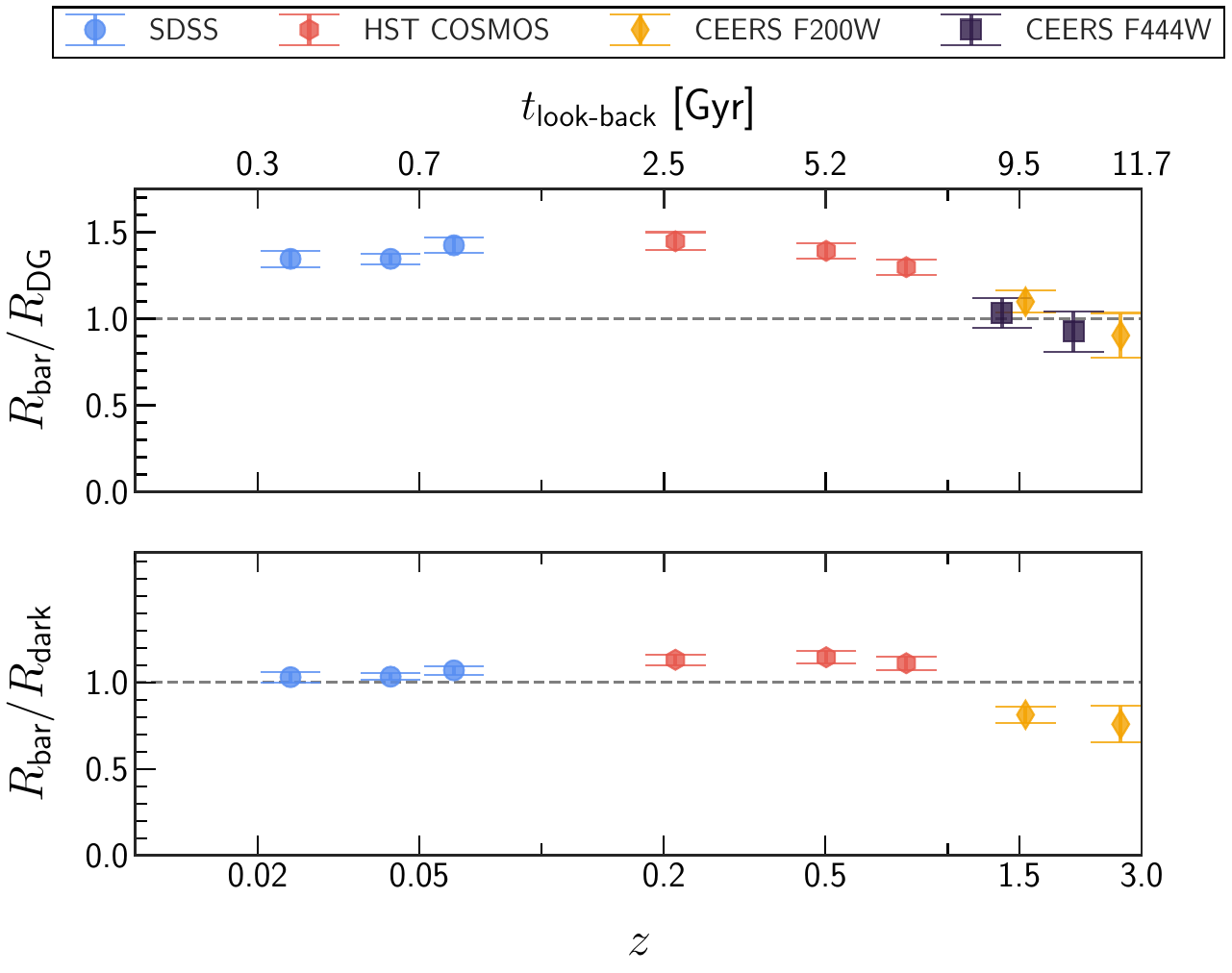}
    \caption{\textit{Redshift evolution of ratio of bar length and dark gap extent:} the average values of ratios, $R_{\rm bar}/R_{\rm DG}$ (top panel) and $R_{\rm bar}/R_{\rm dark}$ (bottom panel), at different redshift bins, shown for our selected sample of barred galaxies from the SDSS, HST COSMOS, and JWST surveys (see the legend). The error bar denotes $\sigma/\sqrt{N}$ spread across each redshift bins, where $N$ is the total number of galaxies in each redshift bin. The axis on top represents the corresponding look-back time, $t_{\rm look-back}$ (in units of Gyr). The horizontal black dashed lines denote $R_{\rm bar}/R_{\rm DG} =1$ (top panel) and $R_{\rm bar}/R_{\rm dark} = 1$ (bottom panel), respectively.}
    \label{fig:rbar_extents_z}
 \end{figure}

\begin{figure}
    \centering
    \includegraphics[width=1\linewidth]{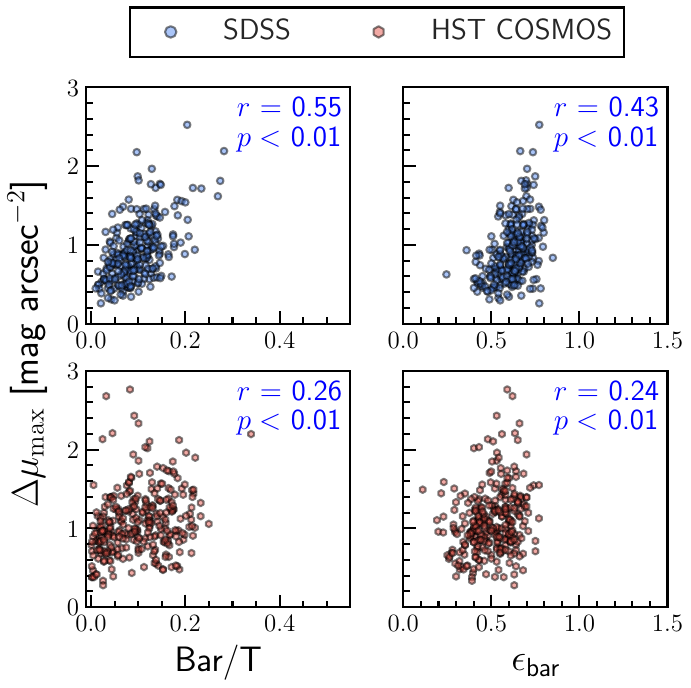}
    \caption{Correlations between the strength of dark gap ($\muMax$) with bar-to-total light ratio Bar/T (left column) and bar ellipticity $\epsilon_{\rm bar}$ (right column), for our  selected barred samples from SDSS (top row) and HST COSMOS (bottom row) surveys is shown here. In each case, the Pearson correlation coefficient, $r$ and the $p$-value is calculated, and the corresponding value is quoted in blue. Both Bar/T and $\epsilon_{\rm bar}$ are found to show a weak ($r \lesssim 0.5$) but statistically significant ($p<0.01$) correlation with $\muMax$.}
    \label{fig:corr3_darkgapstrength_epsBar}
\end{figure}

In the previous section, we investigated how the strength and extent of dark gaps evolve with redshift. Here, we examine whether these properties of dark gaps are correlated with photometric properties of bars as well. Below, we mention the methods used to define the bar length (\rbar) in different surveys considered here.

\begin{itemize}
    \item SDSS - The bar length is defined as the semi-major axis of the bar component in the 2D parametric multi-band multi-component image decomposition of the sample \citep[for further details, see][]{Gadotti2009, Gadotti2011}.
    \item HST COSMOS - Isophotes are fitted to each galaxy and the change in the ellipticity and PA profiles is inspected to determine the bar length. The final deprojected bar length (\rbar) is then calculated analytically using the de-projection procedure developed by \cite{Gadotti2007}.  This process requires parameters such as the angle between the bar and the line of nodes, the ellipticity of the bar, and the inclination of the galaxy, for which the results obtained from GALFIT are used \citep[for details, see][]{kim2021}.
    \item JWST CEERS - The projected bar lengths are measured using the semi-major axis average (sma) of  $\Delta$ PA (change in bar position angle) and $e_{\rm peak}$ (ellipticity peaks) that are obtained via visual adjustments in elliptical isophote fitting \citep[for details, see Sec 5.2 in ][]{LeConteetal2026}.
\end{itemize}
\par
Past theoretical efforts revealed that the extent of dark gaps are well correlated with bar length, \rbar \citep[e.g. see][]{Kimetal2016, Aguerrietal2023,GhoshDiMatteo2024,Ghoshetal2024}. In addition, the recent observational study by \citet{Kimetal2025} confirmed a good correlation between the dark gap radius ($R_{\rm DG}$) and the bar length for nearby galaxies. We first show the distribution of the quantities $R_{\rm DG}$/\rbar and $R_{\rm dark}$/\rbar in Fig.~\ref{fig:corr2_rbar_rdg_rDark} (left panel), calculated for the combined (SDSS + HST COSMOS + JWST CEERS) samples. Both the distributions exhibit well-defined peaks at  0.62 and 0.87 respectively. Then, we examine whether any correlation exists in the physical scales (kpc units), between bar length ($R_{\rm{bar}}$) and dark gap extents ($R_{\rm DG}$, $R_{\rm dark}$) for the wide range of redshift as considered in this work. This is shown in Fig.~\ref{fig:corr2_rbar_rdg_rDark} (right panel). To quantify the correlations, we calculated the Pearson correlation coefficient ($r$) and the corresponding $p$-value, for each of these cases. We find that the correlation between $R_{\rm DG}$ and \rbar ($r = 0.71$) is weaker than that between $R_{\rm dark}$ and \rbar, where a relatively strong correlation ($r > 0.75$) is observed across the entire combined redshift range. We mention that the correlations are also found to be statistically significant ($p$-value $<0.01$). This is in well agreement to the numerical study by \cite{Ghoshetal2024} who also found a better correlation between $R_{\rm bar}$ and $R_{\rm dark}$. Therefore, our results support the fact that the dark gaps and the bar evolve in tandem. To investigate it further in detail, we computed the values of $r$, separately for each the bar samples considered here, in Appendix \ref{appen:item_C}. In short, the degree of correlations tend to vary with the choice of survey. One plausible reason behind the variation of the correlation of \rbar and $R_{\rm DG}$ could be the different methods employed to define the bar length (\rbar). To elaborate, often the bar length derived from the ellipticity profiles overestimates the true bar length \citep[for details, see][]{GhoshDiMatteo2024}. In addition, the bar lengths estimated from the photometric decomposition are often found more robust than those derived from the ellipticity profiles. Lastly, we mention that for the JWST CEERS sample, obtaining the de-projected bar length was not possible due to a large uncertainty in the angle of inclination. A systematic investigation of this is beyond the scope of the present study, and will be taken up in a future theoretical work on bar-driven dark gaps.
\par 
Interestingly, \citet{Aguerrietal2023} showed that for a majority (about 90 percent) of their sample of barred galaxies from CALIFA, ESO/MUSE, ESO/NTT observations survey, the ratio $R_{\rm bar}/R_{\rm DG}$ lies above 1.2. In addition, a recent theoretical study by \citet{Ghoshetal2024} showed that in all the simulated bar models they used, the value of $R_{\rm bar}/R_{\rm DG}$ lies well above 1.2 at all times during the bar evolutionary phase. Here, we investigate if our selected barred samples (covering a wide range of redshift) display any dichotomy around the ratio $R_{\rm bar}/R_{\rm DG} =1.2$. In Fig.~\ref{fig:rbar_extents_z}, we show the evolution of the ratios $R_{\rm bar}/R_{\rm DG}$ and $R_{\rm bar}/R_{\rm dark}$ with redshift, for our selected samples of barred galaxies. We find that overall 63 percent of the total bar sample considered here, the ratio $R_{\rm bar}/R_{\rm DG}$ remains well above 1.2 . In particular, about 71 percent of our selected barred galaxies from the SDSS sample satisfies $R_{\rm bar}/R_{\rm DG} > 1.2$ (also see the top panels of Fig.~\ref{fig:corr2_rbar_rdg_rDark}). This fraction reduces to 64 percent for the HST COSMOS sample, to only 26 percent for the CEERS F200W sample, and 23 percent for CEERS F444W sample. We find a similar evolutionary scenario for the values of $R_{\rm bar}/R_{\rm dark}$ with redshift (see bottom panel of Fig.~\ref{fig:rbar_extents_z}). The implications of this trend is discussed in Sec.~\ref{sec:discussion}. 
\par 
Finally, we examine whether the strength of dark gaps are correlated with the bar-to-total luminosity ratio (Bar/T) as well as bar ellipticity. The ellipticity of the bar is defined as $\epsilon_{\rm bar} = 1 - b/a$, where $a$ and $b$ denote the bar semi-major and semi-minor axes, respectively. A stronger bar is often associated with a higher value of Bar/T. Similarly, a higher value of $\epsilon_{\rm bar}$ denotes a stronger bar, while weaker bars correspond to lower values of $\epsilon_{\rm bar}$ \citep[e.g. see][]{Gadotti2009,Kimetal2016}. The corresponding values for SDSS samples are obtained from BUDAA v2.1 fits on the decomposed images of galaxy  \citep[see][]{Gadotti2009}; while those for HST COSMOS are obtained via photometric analysis in GALFIT \citep[see][]{kim2021}. We mention that calculating accurate values of Bar/T and $\epsilon_{\rm bar}$ for the JWST CEERS sample is currently challenging owing to the limited resolution of the images. Therefore, we limit our study only to the SDSS and HST COSMOS samples. The corresponding correlations between dark gap strength ($\muMax$) with  Bar/T and $\epsilon_{\rm bar}$ are shown in Fig.~\ref{fig:corr3_darkgapstrength_epsBar}. As revealed by the respective values of the Pearson correlation coefficient, the correlation between the $\muMax$ and Bar/T as well as $\epsilon_{\rm bar}$ shows a dependence on the sample we choose. To elaborate, for the SDSS sample, the $\muMax$ and  Bar/T are moderately correlated ($r > 0.5$) while for the HST COSMOS sample,  the $\muMax$ and  Bar/T are weakly correlated. A similar trend is also seen for the correlation between  the $\muMax$ and  $\epsilon_{\rm bar}$ (compare top and bottom right panels of Fig.~\ref{fig:corr3_darkgapstrength_epsBar}). This implies that $\muMax$ is also an important parameter for characterizing bars in itself. 

\section{Discussion}
\label{sec:discussion}
Here, we discuss the implications and a few limitations of this work. First, the properties (strength and extent) of dark gaps (and hence, of bar) show a smooth evolutionary scenario over a wide range of redshift ($0.02 \lesssim z < 3$). We further checked that whether any de-projection effect or a specific choice of a photometric band have any influence to bias our findings. As shown in Appendices~\ref{appen:item_A} and ~\ref{appen:item_B}, these two factors do not alter our main results. Therefore, our findings are robust and demonstrates a relative lack of drastic evolution of dark gap properties (strength and extent) with redshift (see Figs.~\ref{fig:muMax_z} and ~\ref{fig:extents_r90_z}). The plausible implication of the redshift evolution of dark gap properties merits further discussion regarding the redshift evolution of bar properties as they are well connected. The bar formation and subsequent growth scenario, as mostly gleaned from numerical simulations, show an initial rapid growth stage, followed by a steady phase with no drastic growth in terms of strength and length \citep[e.g. see][]{Noguchi1987,Gerinetal1990,Sundinetal1993,MiwaandNoguchi1998,Inmaetal2017,Ghoshetal2023a}. During their lifetime, they often undergo a vertical buckling instability which results in weakening of the bar in relatively shorter timescale \citep[e.g. see][]{Combesetal1990,MerrittandSellwood1994,Debattistaetal2004,Martinez-Valpuestaetal2006,Ghoshetal2023b}. In addition, mechanisms which involve weakening of bars include enhancement of central mass concentration, gas inflow in the central region \citep[e.g. see][]{Pfenniger1990,ShenSellwood2004,Athanassoulaetal2005,Bournaudetal2005,HozumiHernqusit2005,Athanassoula2013}, and minor merger with satellites \citep{Ghoshetal2021}. As shown in this work, the barred galaxies from the JWST CEERS sample show a prominent, well-defined dark gap implying that bars have already formed prior to the observed redshifts, and have quickly evolved. This is consistent with the recent study by \cite{kalitaetal2026} who found matured bars at $z \sim 1.5$. In addition, the moderate increment in the overall dark gap properties (both strength and extent) at higher redshift ($1.4 \leq z < 3$) indicates that the bars (in our selected sample) may as well have passed their initial rapid growth phase. 
In the intermediate ($0.2 \leq  z \leq 1$) as well as lower redshift range ($z < 0.05$), the dark gap properties (as well as the bar properties) stay mostly constant, thereby indicating that the bars remain as mostly stable features (at least, in the statistical sense).
\par
Secondly, our  selected barred samples demonstrate a dichotomy around $R_{\rm bar}/ R_{\rm DG} = 1.2$. While $\sim 63 \%$ of our selected barred galaxies lie above $R_{\rm bar}/ R_{\rm DG} = 1.2$, the remaining $\sim 37 \%$ barred galaxies display $R_{\rm bar}/ R_{\rm DG} < 1.2$. Furthermore, the fraction of galaxies that lie above $R_{\rm bar}/ R_{\rm DG} = 1.2$ decreases steadily with increasing redshift. We mention that the sample size for JWST CEERS sample is much smaller than the other two samples considered here. However, if we assume that the 27 (35) galaxies from CEERS F200W (F444W) sample is a \textit{true} representation of the total bar population in that redshift range, then our finding hints towards a dynamical evolution of the ratio $R_{\rm bar}/R_{\rm DG}$ with redshift. In addition, \citet{Aguerrietal2023} showed that about $90 \%$ of their observed galaxies lie above $R_{\rm bar}/ R_{\rm DG} = 1.2$ and also have $R_{\rm CR}/ R_{\rm DG} \sim 1.8$ (thereby linking dark gaps to ultra-harmonic resonances); while the remaining (non-negligible) $\sim 10 \%$ lie below $R_{\rm bar}/ R_{\rm DG} = 1.2$ and have $R_{\rm CR}/ R_{\rm DG} \sim 1$ (thereby linking dark gaps to the radius of co-rotation, $R_{\rm CR}$). The growth rate of the bar (in terms of size) and the dark gap depend critically on the efficiency of angular momentum transport from the bar region to the dark matter halo and the outer disc region \citep[e.g. see][]{DebattistaSellwood2000,SellwoodandDebattista2006,Ghoshetal2023a,Ghoshetal2024}. The detailed evolution of the ratio $R_{\rm bar}/R_{\rm DG}$ in barred galaxies within a realistic cosmological context is worth pursuing, but currently it is beyond the scope of this work and will be taken up in a future study. However, we mention that the reliability of the conclusion about the fraction of galaxies lying above $R_{\rm bar}/ R_{\rm DG} = 1.2$, especially for the JWST sample, critically depend on the point spread function (PSF) in the sense whether JWST has sufficient resolution to distinguish a $20 \%$ difference in spatial scales. In order to investigate it in further detail, we computed the distribution of the ratio $\frac{|R_{\rm DG} - R_{\rm bar}|}{\rm PSF}$ for both CEERS F200W, F444W samples. In other words, if this ratio remains greater than unity, then measurements are reliable. We found that for 26\% and 49\% of our barred samples from CEERS F200W and F444W respectively, the ratio $\frac{|R_{\rm DG} - R_{\rm bar}|}{\rm PSF}$ remains greater than unity. Therefore, the ratio $R_{\rm bar}/ R_{\rm DG}$ lying close to unity could also be due to the limitations of the PSF of JWST.
\par 
Lastly, for our selected bar sample, we found a weak correlation between $\muMax$ and Bar/T ratio as well as bar ellipticity ($\epsilon_{\rm bar}$). This points to a possibility that $\muMax$ can be an important parameter for characterizing strength of bars in itself. So far, the bar strength ($S_{\rm bar}$) is quantified observationally from the amplitude of the $m=2$ Fourier moment \citep[e.g.][]{Aguerrietal1999,Aguerrietal2000} or using a two-dimensional Fast Fourier transform (FFT) method \citep{Garciaetal2017}. However, until now no systematic study exists which investigates the relation (if any) among various ways of quantifying observationally the strength or prominence of bars. 
 
\section{Summary}
\label{sec:summary}

In summary, we carried out a systematic study of the evolution of bar-driven dark gap properties with redshift. To achieve this, we used a sample of $625$ barred galaxies, selected from three different surveys, namely, the SDSS, HST COSMOS, and JWST CEERS. Our selected barred sample spans a wide range of redshift ($0.02 \lesssim z < 3$), thereby enabling us for the first time, to systematically investigate the dark gap properties over such wide range of redshifts. The main findings of this work are listed below:

\begin{enumerate}
\item{The average values of $\muMax$, quantifying the strength of dark gaps, show a weak evolutionary scenario for a wide range of redshift ($0.02 \lesssim z < 3$), with an increasing trend ($\sim 85$ percent) from the higher redshift regime ($z \sim 2.5$) and a slightly declining trend towards the lower redshift range ($z < 0.05$).}

\item {The average values of $R_{\rm DG}$ or $R_{\rm dark}$ (normalised by \rbar), both quantifying the extent of dark gaps, do not show any drastic evolutionary scenario in the aforementioned redshift range ($0.02 \lesssim z < 3$). In particular, they show a moderately decreasing trend ($\sim 28$ percent) from the higher redshift range ($z \geq 1.4$) and an overall constant trend in the entire intermediate ($0.2 \leq  z \leq 1$) and lower redshift regime ($z < 0.05$), thereafter. Our results suggest that bar formation and the initial rapid growth phase occur before $z \sim 3$, followed by a more gradual growth phase toward lower redshifts.}

\item {For our selected barred galaxy samples from the SDSS, HST COSMOS, and JWST CEERS, the calculated bar length ($R_{\rm bar}$; in units of kpc) is found to be strongly correlated ($r > 0.75$) with $R_{\rm dark}$, but shows a moderate correlation ($r = 0.71$) with $R_{\rm DG}$. Hence, we conclude that $R_{\rm dark}$ is indeed a better proxy (than $R_{\rm DG}$) for \rbar, in agreement with the theoretical study by \cite{Ghoshetal2024}. However, the degree of correlations tend to vary with the choice of survey samples.
In addition, a large fraction ($\sim 63$ \%) of our selected barred samples show that the ratio $R_{\rm bar}/R_{\rm DG}$ lies above 1.2, in agreement with previous observational and theoretical studies. Furthermore, the fraction of galaxies that lie above $R_{\rm bar}/ R_{\rm DG} = 1.2$ decreases steadily with increasing redshift. This might have implications for associating the dark gaps to different bar resonances.}

\item{The strength of dark gaps, $\muMax$ shows a weak but statistically significant correlation with other photometric measurements of bar prominence, for example, bar-to-total light ratio (Bar/T) and bar ellipticity ($\epsilon_{\rm bar}$); implying $\muMax$ is also an independent parameter for characterizing bars.}
\end{enumerate}
\par
Since dark gap properties are excellent proxies for measuring bar properties, as demonstrated in recent theoretical studies, our results have a direct implication on determining the overall evolution of bar properties, from high redshift regime ($z \sim 3$) to local Universe ($z \sim 0.02$). Our findings suggest that some bars may have formed well before the observed redshifts considered here, then quickly evolved through a rapid initial growth phase, followed by a largely steady phase at intermediate redshift range, and finally started to weaken (in terms of strength only) towards the lower redshift regime ($z < 0.05$). In addition, it could also be that for some reason, new bars are itself born weaker at lower redshifts; thereby making the gaps weaker on average. Future JWST observations of bars at higher redshift regime would increase the sample size, which in turn, would facilitate us to carry out a statistically more robust determination of average bar properties as well as their evolutionary pathway at these high redshift regimes.

\section*{ACKNOWLEDGEMENTS}
We thank the anonymous referee for valuable suggestions that have improved the paper. S.C. acknowledges support from the INSPIRE-SHE program (Reg. No. 22MS077) provided by the Department of Science and Technology (DST), Govt. of India. S.G. acknowledges funding from the IIT-Indore, through a Young Faculty Research Seed Grant (project: `INSIGHT'; IITI/YFRSG/2024-25/Phase-VII/02). D.A.G. is supported by STFC grant ST/X001075/1. T.K. was supported by Basic Science Research Program through the National Research Foundation of Korea (NRF), funded by the Ministry of Education (RS-2025-25399934). V.C. acknowledges the support provided by ANID through the FONDECYT grant no. 11250723. E.A. acknowledges financial support by the CNES. This work has made the use of SAOImage DS9, an astronomical imaging and data visualization application developed by the Smithsonian Astrophysical Observatory \citep{saods9_2025}. 

\section*{Data Availability}
\noindent The observational data from the SDSS and HST surveys, used in this work, are already publicly available. For the JWST, the specific observations analysed can be accessed via \url{https://doi.org/10.17909/xm8m-tt59}.

\bibliography{my_ref}
\bibliographystyle{mnras}


\newpage
\appendix

\section{Effect of de-projection}
\label{appen:item_A}

\begin{figure*}
    \centering
    \includegraphics[width=0.875\linewidth]{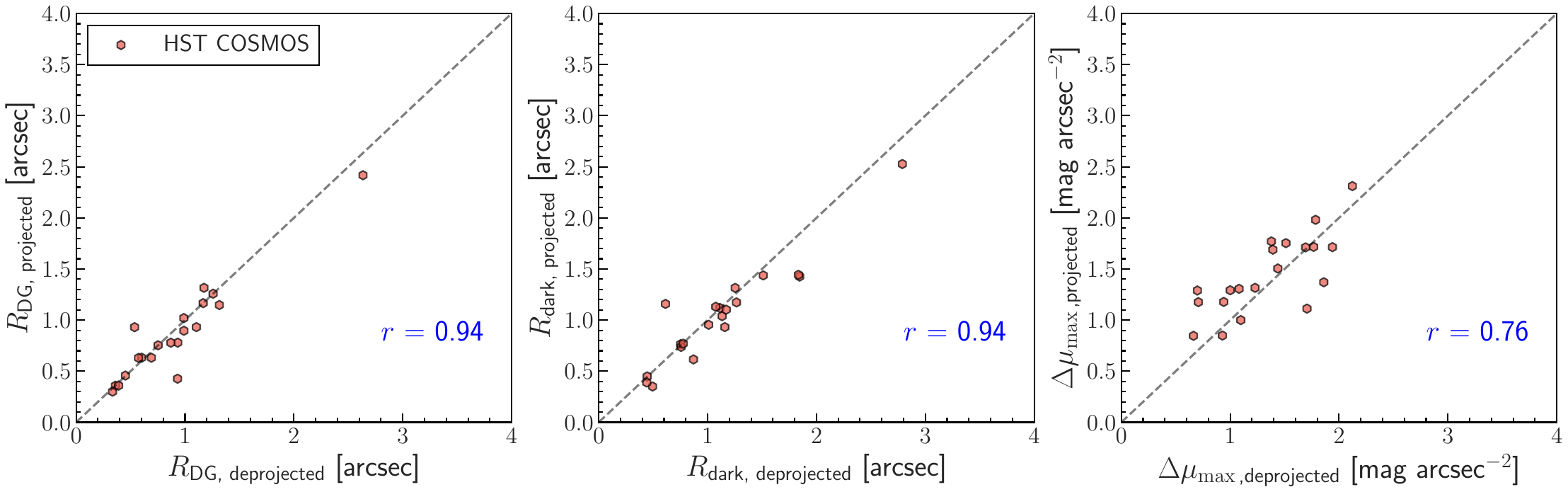}
    \caption{\textit{Effect of de-projection on measuring dark gap properties:} Correlations between dark gap properties, calculated with and without de-projecting the galaxy, for a randomly chosen sample of 20 barred galaxies from our HST COSMOS sample \citep{kim2021}. \textit{Left panel} shows for the $R_{\rm DG}$ measurement, while the \textit{middle panel} and \textit{right panel} show for $R_{\rm dark}$ and $\muMax$, respectively. The grey dashed line in each sub-panel denotes the 1:1 correspondence. In each case, the Pearson correlation coefficient, $r$ is calculated, and the corresponding values are quoted in blue. In each of these measurements, it can be seen that the effect of de-projection is negligible.}
    \label{fig:appdx_depro}
\end{figure*}

As mentioned in Sec.~\ref{sec:redshift_darkgap}, the lack of measurement of the angle of inclination ($i$) for the JWST CEERS sample made it impossible to de-project them and then calculate the radial profiles of surface brightness along the bar major and minor axes. Hence, it is worth checking if this introduces any systematic bias in the calculations of $\muMax$, $R_{\rm DG}$, and $R_{\rm dark}$. In order to investigate this, we make use of our HST COSMOS sample, where measurements of angle of inclination ($i$) are available \citep{kim2021}. First, we randomly choose 20 barred galaxies from the HST COSMOS sample (with varied angle of inclination), and then calculate the values of $\muMax$, $R_{\rm DG}$, and $R_{\rm dark}$ from the HST COSMOS sample \textit{without} de-projecting them. Next, we compare the resulting values with the ones obtained with de-projection of photometric images. We repeated this exercise a few times. The results for one such randomly selected 20 barred galaxies from the HST COSMOS sample are shown in Fig.~\ref{fig:appdx_depro}. The presence of a lesser degree of random scatter around the 1:1 correspondence, together with higher Pearson correlation coefficient ($r > 0.75$), demonstrates that there is no significant systematic error, thereby reinforcing the reliability of our measurement as well as the inferences made based on those values (see Fig.\ref{fig:appdx_depro}).

\section{Effect of band selection}
\label{appen:item_B}

\begin{figure*}
    \centering
    \includegraphics[width=0.85\linewidth]{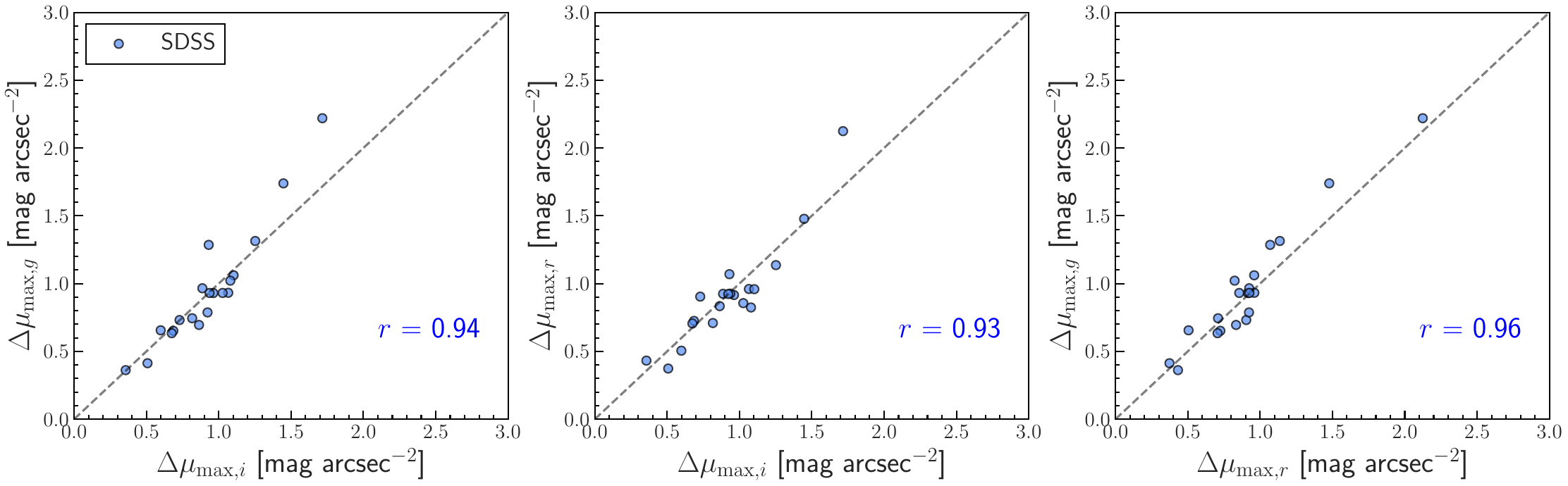}
    \includegraphics[width=0.85\linewidth]{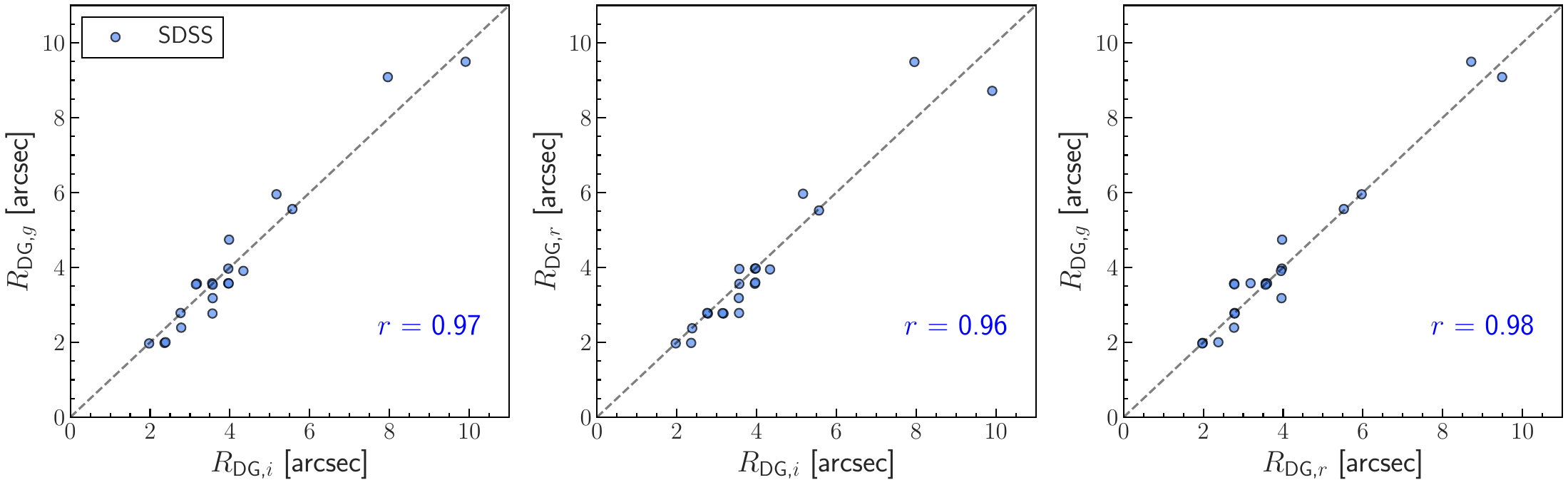}
    \includegraphics[width=0.85\linewidth]{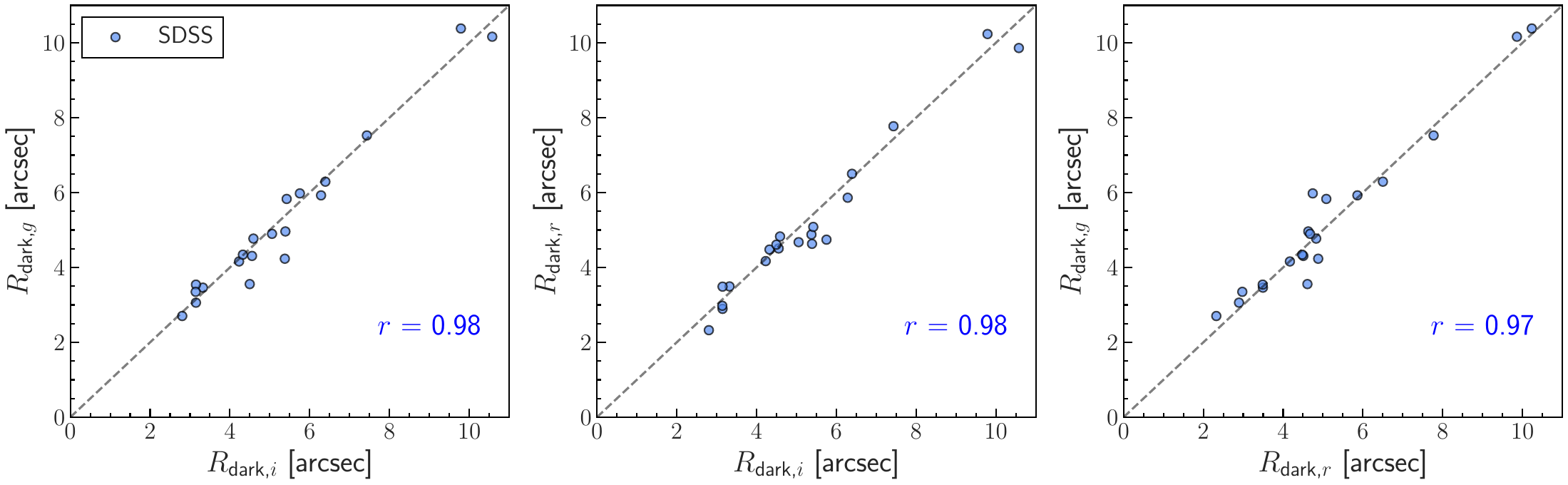}
        \caption{\textit{Effect of photometric band selection on measuring dark gap properties:} Correlations of the dark gap properties measured in three photometric bands $i, \ g,$ and $r$ for 20 randomly chosen galaxies from our SDSS sample \citep{Gadotti2009} are shown. \textit{Top} row shows for $\muMax$, while \textit{middle} and \textit{bottom} rows shows for $R_{\rm DG}$ and $R_{\rm dark}$ respectively, across three different photometric bands. The grey dashed line in each sub-panel denotes the 1:1 correspondence. In each panel, the Pearson correlation coefficient, $r$ is calculated, and the corresponding values are quoted in blue. For each of these measurements, we conclude that the choice of a particular photometric band does not alter the main findings.}
    \label{fig:appdx_bands}
\end{figure*}

 We mention that all the four samples of barred galaxies, chosen from different surveys, are in infrared regime ($0.75 \sim 5 \ \mu$m). While this ensures to obtain a deeper photometric image and a better quantification of dark gap (and hence, bar) properties, it remains to be investigated whether the choice of the band alters the main findings of this work. This is worth investigating, as a recent study by \citet{Menéndez-Delmestre2023}, using a bar sample from the Spitzer Survey of Stellar Structure in Galaxies (S$^4$G) survey, showed that bars appear thinner and longer at bluer bands. To examine that we make use of the SDSS sample of barred galaxies for which photometric images in three different wavelength bands were available. First, we choose 20 randomly selected barred sample and then compute the values of $\muMax$, $R_{\rm DG}$, and $R_{\rm dark}$  in three different bands, namely, $\qty(g, \ r, \ i)$. They are shown in Fig.~\ref{fig:appdx_bands}. The presence of a lesser degree of random scatter around the 1:1 correspondence, together with higher Pearson correlation coefficient ($r > 0.75$), again demonstrates that there is no significant systematic error based on the photometric band chosen for calculating the values of $\muMax$, $R_{\rm DG}$, and $R_{\rm dark}$; thereby not affecting the main findings of this work.

\section{Dependence on the choice of the survey}
\label{appen:item_C}

\begin{figure*}
    \centering
    \includegraphics[width=1\linewidth]{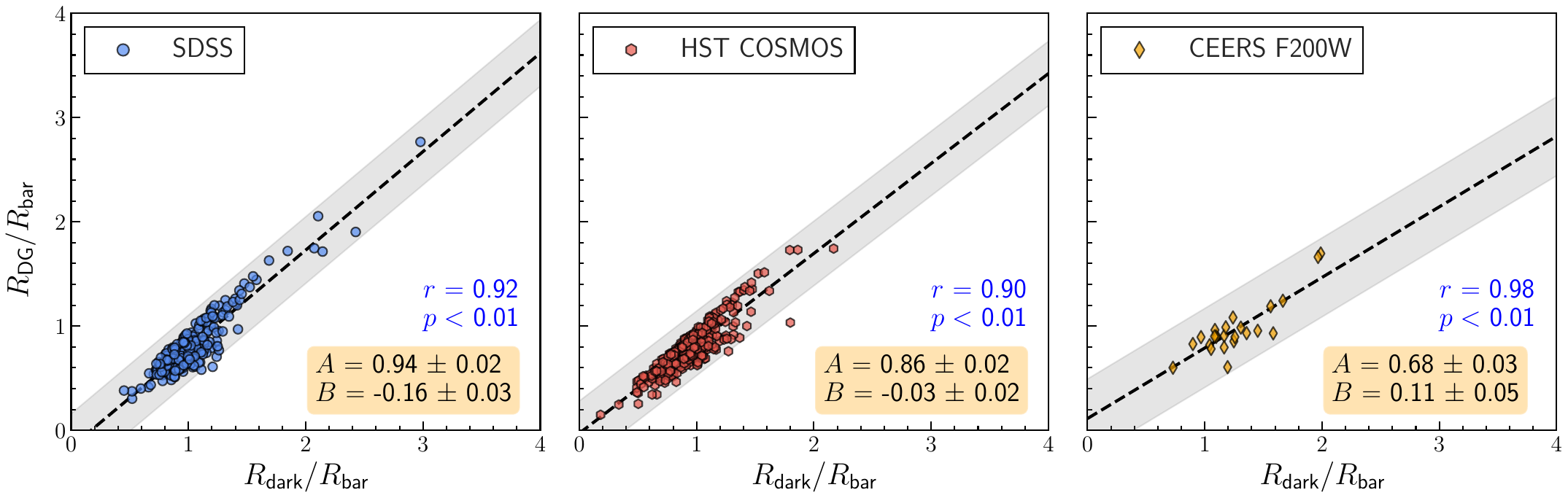}
    \caption{Correlation between the two dark gap extent estimators, $R_{\rm DG}$ and $R_{\rm dark}$ (both normalised by \rbar) calculated separately for our  selected barred samples from each SDSS, HST COSMOS, and JWST (CEERS F200W only) surveys  respectively (see the legends). The two dark gap extent estimators remain strongly correlated (Pearson correlation coefficient, $r > 0.75$). The black dashed line denotes the best-fitting straight line of the form $Y = AX + B$, while the grey shaded region indicates a $3\sigma$ spread around the best fit.}
    \label{fig:corr4_rdg_rDark_samples}
\end{figure*}

\begin{figure*}
    \centering
    \includegraphics[width=1\linewidth]{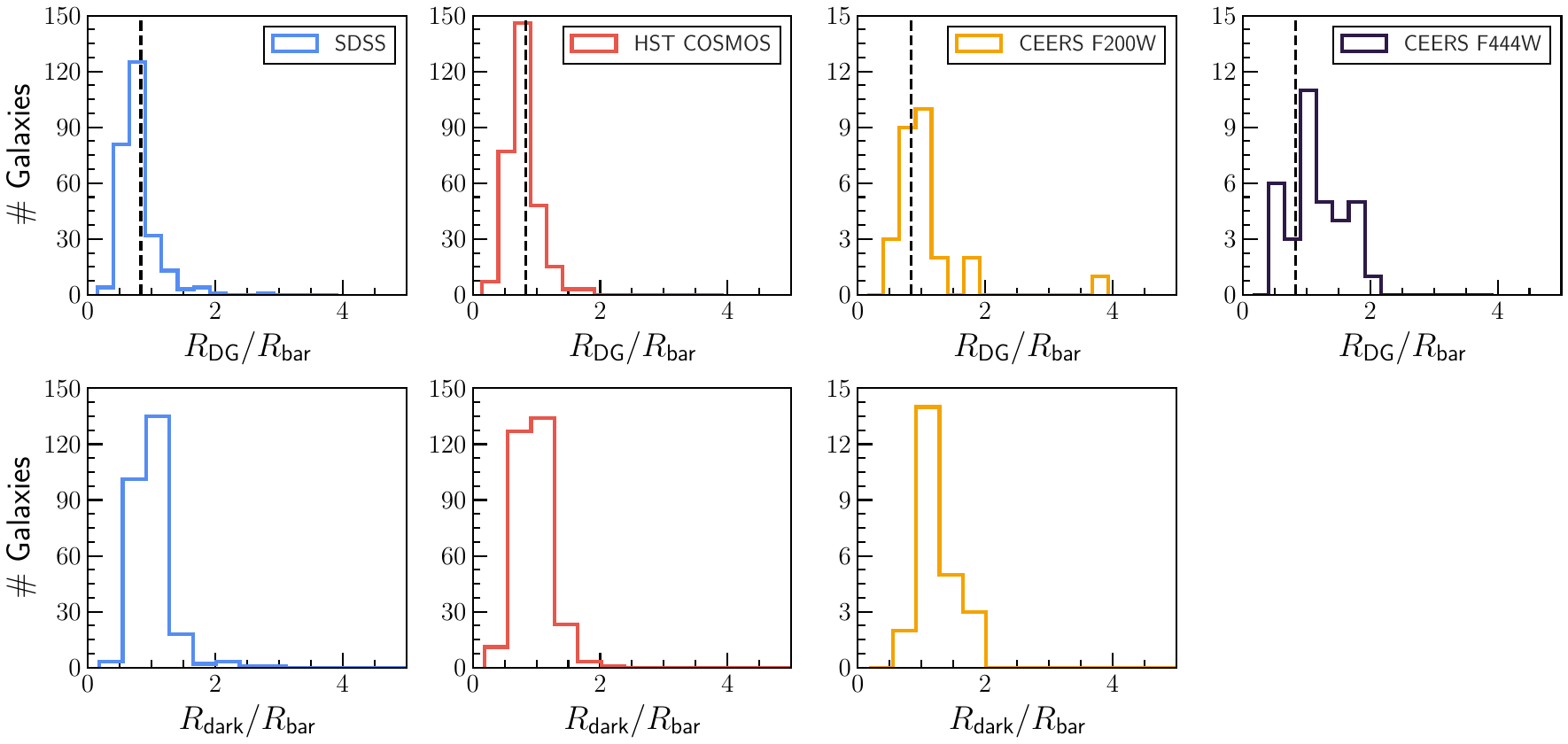}
    \caption{Distributions of $R_{\rm DG}/ R_{\rm bar}$ (top panels) and $R_{\rm dark}/ R_{\rm bar}$ (bottom panels), for our selected sample of barred galaxies from each SDSS, HST COSMOS, and JWST CEERS surveys (see the legends). The vertical black dashed line in the top panels refers to $R_{\rm bar} = 1.2 R_{\rm DG}$. The presence of well defined peaks in the histograms (except for the CEERS F444W sample) provides the evidence of non-zero correlations between the bar length (\rbar) and estimators of dark gap extent ($R_{\rm DG}$ and $R_{\rm dark}$). However, the degree of correlations tend to vary with each survey.}
    \label{fig:hist5_rbar_rdg_rDark_samples}
\end{figure*}


In this work, we make use of barred galaxies, selected from three different surveys, namely, SDSS, HST COSMOS, and JWST CEERS. In Sec.~\ref{sec:correlation_study}, we studied the correlations of the dark gap properties while making no explicit distinction between different surveys (see Figs.~\ref{fig:corr1_rdg_rDark} and \ref{fig:corr2_rbar_rdg_rDark}). Here, we investigate whether the degree of correlation depends on the choice of surveys from which we are selecting the barred samples. These are shown in Figs.~\ref{fig:corr4_rdg_rDark_samples} and \ref{fig:hist5_rbar_rdg_rDark_samples}. 
The strong correlation between $R_{\rm DG}$ and $R_{\rm dark}$ (normalised by \rbar) remains unaffected by the choice of survey, confirming that their linear relation ($Y = AX+B$) is indeed universal. In Fig~\ref{fig:hist5_rbar_rdg_rDark_samples}, we plot the histograms of the ratio $R_{\rm DG}/ R_{\rm bar}$ (in top panels) and $R_{\rm dark}/ R_{\rm bar}$ (bottom panels). All the distributions (except CEERS F444W) exhibit well defined peaks.

Furthermore, we also calculated separately the Pearson correlation coefficient ($r$) between \rbar and $R_{\rm DG}$ (both converted to physical scales; in kpc units) and found the values to be 0.73, 0.68, 0.58, and 0.23 for SDSS, HST COSMOS, CEERS F200W and CEERS F444W, respectively. The corresponding $r$ values between $R_{\rm dark}$ and \rbar (except for the CEERS F444W sample) are reported to be 0.82, 0.71, and 0.52. All the correlations are statistically significant with $p < 0.01$, except for CEERS F444W ($p = 0.19$). Hence, the degree of correlations with bar length is seen to be dependent on the choice of the survey. One plausible reason behind this variation could be the different methods used to compute the bar length. These details will be taken up in a future study.

\bsp
\label{lastpage}

\end{document}